\documentclass[a4paper,11pt]{article}
\usepackage{graphicx}
\usepackage{dcolumn}
\usepackage{bm}
\usepackage{graphicx,amssymb,amsmath,amsfonts,epsfig}
\usepackage[usenames,dvipsnames]{color}

\title{Area (or Entropy) Products in Modified Gravity and Kerr-MOG/CFT Correspondence}
\author{Parthapratim Pradhan\footnote{pppradhan77@gmail.com}\\ 
{\it Department of Physics}\\
{\it Hiralal Mazumdar Memorial College For Women}\\
{Dakshineswar, Kolkata-700035, India}}
\date{}

\begin{document}

\maketitle

\begin{abstract}
We examine the thermodynamic features of \emph{inner} and outer horizons  of 
 modified gravity~(MOG) and its consequences on the holographic duality. 
We derive the thermodynamic product relations for this gravity. We consider both spherically symmetric solutions 
and axisymmetric solutions of MOG. We find that the area  product formula for both cases is \emph{not} 
mass-independent  because they depends on the ADM  mass parameter while in \emph{Einstein gravity} 
this formula is mass-independent~(universal).  We also explicitly verify  the \emph{first law} which is 
fulfilled  at the inner horizon~(IH) as well as at the outer horizon~(OH). We derive thermodynamic products 
and sums for this kind of gravity. 
We  further derive the \emph{Smarr like mass formula} for this kind of black hole~(BH) in MOG.  Moreover, we derive 
the area  bound for both the horizons. Furthermore, we show that the central charges 
of the left and right moving sectors are the same via universal thermodynamic relations.  We also discuss the most 
important result of the \emph{Kerr-MOG/CFT correspondence}. We derive the central charges for Kerr-MOG BH which is 
$c_{L}=12J$ and it is similar to Kerr BH.  We also derive the dimensionless temperature of a extreme Kerr-MOG 
BH which is $T_{L} =  \frac{1}{4\pi} \frac{\alpha+2}{\sqrt{1+\alpha}}$, where $\alpha$ is a MOG parameter. 
This is actually dual CFT temperature of the Frolov-Thorne thermal vacuum state. In the limit $\alpha=0$, we 
find the dimensionless temperature of Kerr BH. Consequently, Cardy formula gives us microscopic entropy for 
extreme Kerr-MOG BH, $S_{micro} = \frac{\alpha+2}{\sqrt{1+\alpha}} \pi J $ for the CFT which is completely in 
agreement with macroscopic Bekenstein-Hawking entropy. Therefore we may conjecture that in the extremal limit 
the Kerr-MOG BH is holographically dual to a chiral 2D CFT with central charge $c_{L}=12J$. Finally, we derive the 
mass-independent area~(or entropy) product relations for regular MOG BH. 
\end{abstract}


\maketitle

\section{Introduction}
Perhaps, a BH is the most fascinating as well as thermal object~\cite{bcw73} of the universe. A thermal object 
in the sense that it has both temperature and entropy~\cite{bk73}. Again this entropy is proportional to the area 
of the event horizon~(EH) and Cauchy horizon~(CH).  They are defined as 
\begin{eqnarray}
{\cal S}_{\pm} &=& \frac{{\cal A}_{\pm}}{4} ~. \label{mg1}
\end{eqnarray}
where, ${\cal S}_{\pm}$ is the so called the Bekenstein-Hawking entropy~(in units in which $G=\hbar=c=k_{B}=1$) and 
${\cal A}_{\pm}$ is the area of both the horizons. Similarly, the temperature is proportional to the surface gravity 
of EH~(${\cal H}^{+}$) and CH~(${\cal H}^{-}$). They are defined as 
\begin{eqnarray}
T_{\pm} &=& \frac{\kappa_{\pm}}{2 \pi}  ~. \label{mg2}
\end{eqnarray}
where $T_{\pm}$ is the so-called the Hawking temperature computed on ${\cal H}^{\pm}$ and
$\kappa_{\pm}$ is defined as the surface gravity of the BH computed on ${\cal H}^{\pm}$.

It is established that in terms of the above quantities, the first law of BH thermodynamics 
for both horizons~(${\cal H}^{\pm}$) becomes
\begin{eqnarray}
dM &=& \pm \frac{{\kappa}_{\pm}} {8\pi} d{\cal A}_{\pm} + \Omega_{\pm} dJ +\Phi_{\pm}dQ ~. \label{mg3}
\end{eqnarray}
where $\Omega_{\pm}=\frac{\partial  M}{\partial J}$ and $\Phi_{\pm} = \frac{\partial M}{\partial Q}$.

In recent times, the area~(or entropy) product formula of ${\cal H}^{\pm}$ has been a fascinating topic of research 
due to the seminal work of Ansorg and  Hennig~\cite{ah09}. In their work, the authors derived that for a Kerr-Newman~(KN) 
class of BH in the Einstein-Maxwell~(EM) gravity the area~(or entropy) product formula of ${\cal H}^{\pm}$ becomes 
\begin{eqnarray}
{\cal A}_{+} {\cal A}_{-} &=& 64 \pi^2\left(J^2+\frac{Q^4}{4}\right) ~.\label{mg5}
\end{eqnarray}
This  is a \emph{universal product relation} in the sense that it is independent of the Arnowitt-Deser-Misner~(ADM) 
mass parameter [See also ~\cite{mv13,plb,pp14,ppgrg,ijmpd,pp15,jetpl,loga,sen}].  

On the other hand, Cveti\v{c} et al.~\cite{cgp11} proposed that for BHs in $D=4$ and $D\geq 4$ the area~(or entropy)
product formulas obey the quantization formula 
\begin{eqnarray}
{\cal A}_{+} {\cal A}_{-} &=& \left(8\pi {\ell _{pl}}^2\right)^2 N , \,\,\, N\in {\mathbb{N}}
  ~.\label{mg6}
\end{eqnarray}
where $\ell _{pl}$ is the Planck length~\cite{finn,castro12,det12,val13,chen12}.

It should be emphasized that one of the major achievements in string theory is that of holographic duality which 
connects quantum theory of gravity to the quantum field theory without any nomenclature of Einstein's general thory of 
gravity~\cite{juan}. 
For the extreme class of BH in string theory, Strominger and Vafa~\cite{sv} have successfully given the idea of the 
microscopic origin of the Bekenstein-Hawking entropy but up to date no idea has been found in the literature for the 
non-extreme class of BH. 
It may be noted that a Kerr BH is dual to a CFT for $AdS_{3}$ BH, which was introduced first by Brown and 
Henneaux~\cite{brown}. Guica et al. ~\cite{guica}~(See also ~\cite{castro10}) first demonstrated the Kerr/CFT 
correspondence by using the near-horizon limiting procedure. They in fact computed the central charges for the 
Kerr BH by using the asymptotic symmetry group~(ASG)\footnote{The ASG is a one type of conformal group.} method 
by imposing some boundary 
conditions. They also proved that by using some boundary conditions, the extreme Kerr BH has a 
feature which is dual to a chiral CFT. When one takes the extreme limit for Kerr BH, one obtains the Frolov-Thorne 
vacuum~\cite{frolov89} state for extreme Kerr BH which produces a thermal state with a temperature $T_{L}=\frac{1}{2\pi}$. 
They also microscopically computed the entropy for an extremal Kerr BH by using the Cardy formula and proved that it generates 
the macroscopic Bekenstein-Hawking entropy formula. We derive this result for the Kerr-MOG BH by
\emph{using thermodynamic procedure.}

However, in this work we wish to examine the area~(or entropy) product relations, broadly speaking we
would like to investigate the thermodynamic properties of \emph{inner} and outer horizons of 
scalar-tensor-vector gravity~(STVG) or MOG~\cite{jcapmf}. 
We have considered both the spherically symmetric solution and the axisymmetric solution of MOG. We show that 
the area~(or entropy) product formula for both the situations is \emph{mass-dependent}  while in Einstein gravity~(EG)
this formula is \emph{mass-independent}.
We also explicitly verify  the \emph{first law} of BH mechanics which is completely in agreement with both the 
inner horizon~(IH) as well as the outer horizon~(OH). We further derive the other thermodynamic relations like 
thermodynamic products and sums. 
Moreover, we derive the 
\emph{Smarr type of mass formula} for this class of BH in MOG. Also, we derive  the area~(or entropy) bound for both the 
horizons. Finally, we show that the central charges of the  left and right moving sectors of the dual CFT  
in MOG/CFT correspondence are the same by using universal thermodynamic relations. 

It should be noted the argument made in~\cite{chen12,chen13},``the area product being mass-independent is equivalent 
to the relation $T_{+}{\cal S}_{+}= T_{-}{\cal S}_{-}$'', where $T_{\pm}$ and ${\cal S}_{\pm}$ are the Hawking temperature 
and the entropies of ${\cal H}^{\pm}$. We show that this argument is violated in case of a Kerr-MOG BH. Because for a
Kerr-MOG BH, the relation $T_{+}{\cal S}_{+}= -T_{-}{\cal S}_{-}$ is satisfied while the area product is 
\emph{not} mass-independent. This is one of the key predictions for a \emph{Kerr-MOG BH}.

Furthermore, we have derived the central charges of the left moving sectors and right moving sectors of dual 
CFT in Kerr-MOG/CFT correspondence. We found that the central charges are the same in both sectors 
i.e. $c_{L}=c_{R}=12J$. We also computed the dimensionless temperature of microscopic CFT, which is completely 
in agreement with the ones computed from 
hidden conformal symmetry. In the extremal limit, we determine the temperature of Frolov-Thorne vacuum state 
which is the so-called dimensionless temperature $T_{L}=\frac{1}{4\pi} \frac{\alpha+2}{\sqrt{1+\alpha}}$. 
When one takes the limit $\alpha=0$, one obtains the dimensionless temperature of the Kerr BH.  Using the 
Cardy formula, we compute microscopic entropy for an extreme Kerr-MOG 
BH, $S_{micro} = \frac{\alpha+2}{\sqrt{1+\alpha}} \pi J $ for the CFT,  which is exactly in agreement 
with the macroscopic Bekenstein-Hawking entropy. Thus we can conjecture that an extreme Kerr-MOG
BH is holographically dual to a chiral 2D CFT with the central charge $c_{L}=12J$.

It is well known that general relativity is the most successful and well examined theory of gravity. However, the STVG
is a modification of laws of gravitation on a length scale where Newtonian gravity or general gravity have not been 
explicitly examined. One such type of gravity is called MOG. This framework correctly explains the observations 
of the solar system~\cite{jcapmf}, the rotation curves of the cluster of galaxies and the dynamics of galaxies clusters. 
There has been no needs to the idea of dark matter~\cite{mf1,mf2,mf3,mf4}. 

One characterisic is that the STVG is a formulation of MOG where the fields are scalar fields and massive vector fields, and 
it can be used to describe the growth of the structure and the power spectrum of matter and the acoustical power spectrum of 
the cosmic microwave background~(CMB) data.

Now  we shortly review the modified action for the STVG~\cite{jcapmf,mf7} which is given by 
\begin{eqnarray}
{\cal I} &=& {\cal I}_{G} +{\cal I}_{V} +{\cal I}_{S} +{\cal I}_{M},  ~ \label{mg7}
\end{eqnarray}
where  ${\cal I}_{G}$ is the Einstein-Hilbert action for gravity, ${\cal I}_{V}$ is the action for massive 
vector field $\phi_{a}$, ${\cal I}_{S}$ is the  action for scalar fields and ${\cal I}_{M}$ is the action 
for pressure less matter. They are defined as 
\begin{eqnarray}
{\cal I}_{G} &=& \frac{1}{16\pi G}\int (R+2\Lambda) \sqrt{-g}\, d^4x,
\end{eqnarray}
\begin{eqnarray}
{\cal I}_{V} &=& -\frac{1}{4\pi} \int \biggl[{\cal K}+V(\phi_{a})\biggr] \sqrt{-g}\, d^4x,
\end{eqnarray}
\begin{equation}
\label{Sac} {\cal I}_{S} = \int\frac{1}{G}
\biggl[\frac{1}{2}g^{ab}\biggl(\frac{\nabla_{a}
G\nabla_{b}G}{G^2}+\frac{\nabla_{a}\mu\nabla_{b}\mu}{\mu^2}\biggr)
$$ $$
-\frac{V_{G}(G)}{G^2}-\frac{V_{\mu}(\mu)}{G^2}\biggr]\sqrt{-g}\, d^4x ~. \label{mg8}
\end{equation}
and 
\begin{eqnarray}
{\cal I}_{M} &=&-\int \biggl[\rho \sqrt{u^{a}u_{a}}+{\cal Q} u^{a}\phi_{a}\biggr] \sqrt{-g}\, d^4x+{\cal J}^{a}\phi_{a} 
\end{eqnarray}
where $R=g^{ab}R_{ab}$, $g=det(g_{ab})$, $\nabla_{a}$ is the covariant derivative corresponds to the metric $g_{ab}$.
The potential, $V(\phi_{a})$ indicates that the potentials are associated with the vector field $\phi_{a}$
and $V_{G}(G)$, $V_{\mu}(\mu)$ denote the potentials with respect to the scalar fields, $G$ and $\mu$, respectively
\footnote{ The MOG theory~\cite{mf1} admits the parameters $G$, $\omega$ and $\mu$. Where $G$ is the Gravitational 
constant, $\omega$ is the coupling constant and $\mu$ is the mass of the massive vector field which is vary with 
space \& time. The coupling constant $\omega$ in the STVG action in~\cite{mf1} is the scalar field and it should be 
treated as a constant value thus hereafter we assume throughout the work $\omega=1$.}.

We have also used the value $c=1$. Finally, the kinetic term with respect to the field $\phi_{a}$ is defined by 
\begin{eqnarray}
 {\cal K} &=&\frac{1}{4}B^{ab}B_{ab}~. \label{mg9}
\end{eqnarray}
where $B_{ab}=\partial_{a}\phi_{b}-\partial_{b}\phi_{a}$.

Now the field equation for the STVG~\cite{jcapmf} is
\begin{eqnarray}
G_{ab}-\Lambda g_{ab}+{\cal Q}_{ab} &=& -8\pi G {\cal T}_{ab}  ~. \label{mg10}
\end{eqnarray}
where ${\cal Q}_{ab}=G\left(\nabla^{a}\nabla_{a}\Theta g_{ab}-\nabla_{a}\nabla_{b}\right)$ and 
$G_{ab}=R_{ab}-\frac{1}{2}g_{ab}R$, $\Lambda$ is the cosmological constant and $\Theta=\frac{1}{G}$. We do 
not write the explicit field equations for $B^{ab}$ because it can be found in~\cite{mf7}. Now we should
define the covariant current density as 
\begin{eqnarray}
 {\cal J}^{a} &=&\kappa {\cal T}^{mab} u_{b}~. \label{mg11}
\end{eqnarray}
where ${\cal T}^{mab}$ is the energy-momentum tensor for matter, $\kappa=\sqrt{\alpha G_{N}}$, $\alpha=\frac{G-G_{N}}{G_{N}}$ 
is the scalar field, $G_{N}$ is the Newtonian constant and $u^{a}$ is the four velocity. 

The perfect fluid energy-momentum tensor for matter is defined as 
\begin{eqnarray}
 {\cal T}^{mab} &=& (\rho_{m}+p_{m})u^{a}u^{b}-p_{m}g^{ab} ~. \label{mg12}
\end{eqnarray}
where $\rho_{m}$ and $p_{m}$ are correspond to the density and pressure of matter respectively. Using the normalization 
condition of the four velocity and with the help of Eq.(\ref{mg11}), and Eq. (\ref{mg12}) one obtains
\begin{eqnarray}
 {\cal J}^{a} &=& \kappa \rho_{m} u^{b}~. \label{mg13}
\end{eqnarray}
Whereas the gravitational source charge is defined as
\begin{eqnarray}
 {\cal Q} &=& \kappa \int {\cal J}^{0}(x)\, d^3x ~. \label{mg14}
\end{eqnarray}
The values ${\cal Q}=\sqrt{\alpha G_{N}}M$ and $G=G_{N} (1+\alpha)$ are derived from the 
weak field approximation~\footnote{The perturbation of the metric around the Minkowski metric $\eta_{ab}$ can be 
written in the form $g_{ab}=\eta_{ab}+\lambda h_{ab}$} of the STVG field equations.

One could study the geodesic motion by using the geodesic equation of a test particle which should read, for the 
time-like particle 
\begin{eqnarray}
\frac{d^2x^{a}}{d\tau^2}+\Gamma^{a}_{bc} \frac{dx^{b}}{d\tau}\frac{dx^{c}}{d\tau}=\frac{q}{m} B_{d}^{a}\frac{dx^{d}}{d\tau}
~. \label{mg14.1}
\end{eqnarray}
and for the light-like particle
\begin{eqnarray}
 \frac{d^2x^{a}}{d\lambda^2}+\Gamma^{a}_{bc} \frac{dx^{b}}{d\lambda}\frac{dx^{c}}{d\lambda}=0
~. \label{mg14.2}
\end{eqnarray}
where $\lambda$ is an affine parameter and $\Gamma^{a}_{bc}$ denotes the Christoffel symbols.

The parameters $m$ and $q=\sqrt{\alpha G_{N}}m$ denote test particle mass and gravitational charge 
respectively. Let us choose the potential for the massive vector field, $\phi_{a}$ of the form 
\begin{eqnarray}
 V(\phi_{a})=-\frac{1}{2} \mu^2 \phi^{a}\phi_{a}, ~\label{mg14.3}
\end{eqnarray}
where $\mu$ is the mass of massive vector field and $\partial_{b}\phi^{b}=0$, and 
$\phi_{a}=(\phi_{0}, \phi_{i}) (i=1,2,3)$. Therefore the radial source free 
static, spherically symmetric solutions of $\phi_{0}$ could be obtained from 
the following equation 
\begin{eqnarray}
 \frac{d^2\phi_{0}}{dr^2}+\frac{2}{r} \frac{d\phi_{0}}{dr}-\mu^2\phi_{0} &=& 0
~. \label{mg14.4}
\end{eqnarray}
The solution of the above equation is $\phi_{0}(r)=-{\cal Q} \frac{e^{(-\mu r)}}{r}$~\footnote{The main interesting 
feature of MOG is that the weak field acceleration law is attractive and repulsive. The Yukawa contribution  due to 
a spin 1~(graviton) is a repulsive force, while the scalar spin 0~(graviton) described by the scalar field $G$ is 
an attractive~\cite{mf1} one.} and the gravitational charge ${\cal Q}=\sqrt{\alpha G_{N}}M$, and $M$ 
is the mass of the source particle.
For matter-free MOG, one has to set the energy momentum tensor equal to zero as well as the cosmological constant  
and then one obtains the energy momentum tensor for the vector field $\phi_{a}$~\cite{mf7} which is briefly discussed 
in the following section.

The manuscript is organized as follows. In Sec. 2, we study the thermodynamic properties of a  
static, Spherically symmetric MOG BH. In Sec. 3, we analyze the thermodynamic properties of the Kerr-MOG BH. 
In Sec. 4, we give most important results of the Kerr-MOG/CFT correspondence. Sec. 5 is devoted to studying 
the thermodynamic properties of a regular MOG BH. Finally, in Sec. 6, we summarize 
the results.

\section{Area Product formula in a static, Spherically symmetric MOG BH}

To derive the static, spherically symmetric solution of a MOG BH we shall use the modified Einstein's field equations 
which can be written as~\cite{mf6} 
\begin{eqnarray}
G_{ab} &=&-8\pi G {\cal T}_{\phi ab}  ~. \label{mg15}
\end{eqnarray}
Since we are working with matter-free MOG field equations, the the energy-momentum tensor of the matter
sector, ${\cal T}_{mab}=0$ and the energy-momentum tensor for the massive vector field, $\phi_{a}$ read 
\begin{eqnarray}
{\cal T}_{\phi ab} &=& -\frac{1}{4\pi} \left(B_{a}^{c} B_{bc}-\frac{1}{4} g_{ab} B^{ef}B_{ef} \right)  ~. \label{mg16}
\end{eqnarray}
where $B_{ab}$ is previously defined and $\phi_{a}$ is the vector field with source charge, 
${\cal Q}=\sqrt{\alpha G_{N}}M$.  We also require other vacuum field equations:
\begin{eqnarray}
 \nabla_{b}B^{ab} &=& \frac{1}{\sqrt{-g}} \partial_{a}\left(\sqrt{-g}B^{ab}\right)= 0~, \label{mg17}
\end{eqnarray}
and
\begin{eqnarray}
\nabla_{a}B_{bc}+\nabla_{b}B_{ca}+\nabla_{c}B_{ab} &=& 0~. \label{mg18}
\end{eqnarray}
where $\nabla_{a}$ denotes the covariant derivative with respect to the metric tensor $g_{ab}$. Now, we assume the static, 
spherically symmetric metric whose form is given by
\begin{eqnarray}
ds^2=e^{\eta} dt^2-e^{\mu} dr^2- r^{2} d\Omega^2 ~. \label{mg19}
\end{eqnarray}
where $d\Omega^2=d\theta^2+\sin^2 \theta d\phi^2$. For the static solution, $\phi_{0} \neq 0$ and 
$\phi_{1}=\phi_{2}=\phi_{3}=0$. Also from Eq. \ref{mg17}, one obtains
\begin{eqnarray}
 \partial_{r}\left(\sqrt{-g}B^{0r}\right)&=&-\sin\theta 
 \partial_{r}\left(e^{\left[-\frac{(\mu+\eta)}{2}\right]}r^2\phi_{0}'\right) = 0~, \label{mg20}
\end{eqnarray}
where $\phi_{0}'=\partial_{r}\phi_{0}$. After integration, one finds  
\begin{eqnarray}
\phi_{0}' &=& e^{\left[\frac{(\mu+\eta)}{2}\right]}\frac{Q}{r^2} ~, \label{mg21}
\end{eqnarray}
where $Q$ is the gravitational source charge of $B_{ab}$. The components of the  
energy-momentum tensor for the massive vector field are
\begin{eqnarray}
{\cal T}_{\phi 0}^{0} &=& {\cal T}_{\phi 1}^{1}=-{\cal T}_{\phi 2}^{2}=-{\cal T}_{\phi 3}^{3}=
\frac{1}{2} e^{(-\mu-\eta)} \left(\phi_{0}'\right)^2=\frac{Q^2}{8\pi r^4} 
~. \label{mg22}
\end{eqnarray}
Now let us put $\lambda=e^{\eta}$ and solving Eqs. (\ref{mg15}), one can obtain $\eta'=-\mu'$ and 
\begin{eqnarray}
 \lambda+r\lambda' &=& 1-\frac{GQ^2}{r^2}, \\
 r\lambda &=& r+\frac{GQ^2}{r}-2GM ~. \label{mg23}
\end{eqnarray}
where $2GM$ is an integration constant. Substituting these values in Eq. (\ref{mg19}), one obtains the gravitational 
field metric 
\begin{eqnarray}
ds^2=\left[1-\frac{2GM}{r}+\frac{GQ^2}{r^2}\right]dt^2-
\frac{dr^2}{\left[1-\frac{2GM}{r}+\frac{GQ^2}{r^2}\right]}-r^{2}d\Omega^2 ~. \label{mg24}
\end{eqnarray}
where $G=G_{N} (1+\alpha)$. Interestingly, the form of this metric is similar to the static, spherically symmetric 
Reissner-Nordstr\"{o}m solution of a charged body and  the value of ${\cal Q}>0$. 

Now by postulating that this independent charge ${\cal Q}$ is proportional to the mass of the source particle i.e. 
${\cal Q}=\kappa M$, we also have $\kappa=\pm \sqrt{\alpha G_{N}}$, of which we have previously defined the value of 
$\alpha$. For physical stable stars, galaxies etc. and to maintain the repulsive gravitational Yukawa force we choose 
the value of independent charge to be positive i.e. ${\cal Q}>0$. Therefore the metric of Eq. (\ref{mg24}) becomes 
\begin{eqnarray}
ds^2=\left[1-\frac{2GM}{r}+\frac{\alpha GG_{N}M^2}{r^2}\right]dt^2-
\frac{dr^2}{\left[1-\frac{2GM}{r}+\frac{\alpha G G_{N}M^2}{r^2}\right]}-r^{2}d\Omega^2 ~. \label{mg25}
\end{eqnarray}
After substituting the value of $G=G_{N} (1+\alpha)$, one obtains the static, spherically symmetric MOG 
BH~\cite{mf5} 
$$
ds^2=\left[1-\frac{2G_{N}(1+\alpha)M}{r}+\frac{G_{N}^2M^2\alpha(1+\alpha)}{r^2}\right]dt^2-
$$
\begin{eqnarray}
\frac{dr^2}{\left[1-\frac{2G_{N}(1+\alpha)M}{r}+\frac{G_{N}^2M^2\alpha(1+\alpha)}{r^2}\right]}
- r^{2}d\Omega^2 ~. \label{mg1.1}
\end{eqnarray}
where $G_{N}$ is the modified Newtonian constant which is related to the Newtonian constant via the relation 
$G=G_{N} (1+\alpha)$ and the modified charge parameter is ${\cal Q}=\sqrt{\alpha G_{N}}M$, where $\alpha$ is 
a free parameter.  The above metric can be obtained 
by putting these values in the usual Reissner-Nordstr\"{o}m BH solution.

The BH has both EH and CH situated at 
\begin{eqnarray}
 r_{\pm} &=& G_{N} M \left(1+\alpha \pm \sqrt{1+\alpha} \right) ~. \label{mg1.2}
\end{eqnarray}
$r_{+}$ is called EH and $r_{-}$ is called CH. The MOG BH does not posess a naked singularity. 
The  area of both horizons (${\cal H}^\pm$) for the MOG BH is
\begin{eqnarray}
{\cal A}_{\pm} &=& 4\pi r_{\pm}^2 =4\pi (1+\alpha) M^2 G_{N}^2 
\left[\alpha+2\left(1+\alpha \pm \sqrt{1+\alpha}\right) \right] ~.\label{mg1.3}
\end{eqnarray}  
The  Bekenstein-Hawking entropy of ${\cal H}^\pm$ (in units in which $\hbar=c=1$) should read
\begin{eqnarray}
{\cal S}_{\pm} &=& \frac{{\cal A}_{\pm}}{4G} =\pi G_{N}  M^2 
\left[\alpha+2\left(1+\alpha \pm \sqrt{1+\alpha}\right) \right] ~.\label{mg1.4} 
\end{eqnarray}

The Hawking temperature of ${\cal H}^\pm$ reads 
\begin{eqnarray}
T_{\pm} &=&\frac{{\kappa}_{\pm}}{2\pi} =\pm \frac{1}{2\pi G_{N} M \sqrt{1+\alpha}\left(1\pm \sqrt{1+\alpha} \right)^2}
~.\label{mg1.5} 
\end{eqnarray}
where ${\kappa}_{\pm}$ is called surface gravity of ${\cal H}^\pm$. 

The Smarr formula is derived to be
\begin{eqnarray}
 M^2 &=&  \frac{{\cal A}_{\pm}}{4\pi (1+\alpha) G_{N}^2 
\left[\alpha+2\left(1+\alpha \pm \sqrt{1+\alpha}\right) \right]}    ~.\label{sm}
\end{eqnarray}  
and the \emph{first law} is satisfied to be 
\begin{eqnarray}
\pm dM &=& \sqrt{1+\alpha} T_{\pm} d{\cal S}_{\pm}  ~. \label{law1}
\end{eqnarray}

The specific heat for a MOG BH is given by
\begin{eqnarray}
C_{\pm} &=& \frac{\partial M}{\partial T_{\pm}}=-2\pi G_{N} M^2 \sqrt{1+\alpha} \left(1\pm \sqrt{1+\alpha} \right)
\left(\sqrt{1+\alpha} \pm 1 \right) .~\label{mg1.6}
\end{eqnarray}

The Komar energy is calculated to be 
\begin{eqnarray}
E_{\pm} &=& 2 T_{\pm} {\cal S}_{\pm}=\frac{\pi M^3 G_{N}^3 \alpha^2 (1+\alpha)^\frac{3}{2}}
{\left(1\pm \sqrt{1+\alpha} \right) \left(\sqrt{1+\alpha} \pm 1 \right)} .~\label{mg1.7}
\end{eqnarray}

Finally, the Gibbs free energy is given by 
\begin{eqnarray}
G_{\pm} &=& M- T_{\pm} {\cal S}_{\pm}=M \left[1-\frac{\pi M^2 G_{N}^3 \alpha^2 (1+\alpha)^\frac{3}{2}}
{2 \left(1\pm \sqrt{1+\alpha} \right) \left(\sqrt{1+\alpha} \pm 1 \right)} \right] .~\label{mg1.8}
\end{eqnarray}

\begin{figure}
\begin{center}
{\includegraphics[width=0.45\textwidth]{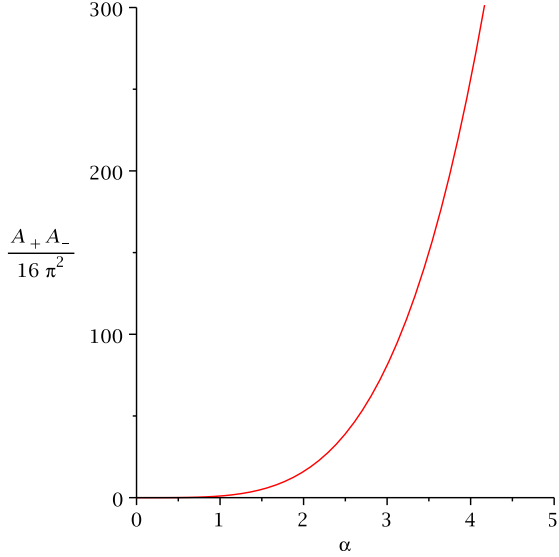}}
\end{center}
\caption{The figure shows the variation of area product of ${\cal H}^{\pm}$ with free parameter $\alpha$ for 
spherically symmetric MOG BH with $M=G_{N}=1$. \label{mogs}}
\end{figure}

The main interest here is to examine the area product formula of  ${\cal H}^\pm$ whether it is mass independent 
or not. Thus, the product is computed to be 
\begin{eqnarray}
{\cal A}_{+} {\cal A}_{-} &=& 16 \pi^2 \alpha^2 (1+\alpha)^2 M^4 G_{N}^4 .~\label{mg1.9}
\end{eqnarray}
It implies that the area (or entropy ) product formula does depend on the mass parameter thus the area (or entropy) 
product formula for the MOG BH in spherically symmetric cases is \emph{not} universal. It is also clearly evident that 
all the other thermodynamic products are \emph{mass dependent}. In Fig. \ref{mogs}, we show how the area product 
varies with the free parameter in the case of a spherically symmetric MOG BH.

Now the irreducible mass product of ${\cal H}^{\pm}$ for the MOG BH is 
\begin{eqnarray}
M_{irr,+}\, M_{irr,-} &=& \frac{M^2 G_{N}^2 \alpha(1+\alpha) }{4} ~. \label{mg1.101}
\end{eqnarray}
where the  irreducible mass $M_{irr, \pm}$ of ${\cal H}^{\pm}$ is defined by
\begin{eqnarray}
M_{irr, \pm} &=& \sqrt{\frac{{\cal A}_{\pm}}{16\pi}} ~. \label{mg1.102}
\end{eqnarray}

The other relevant thermodynamic relations are
\begin{eqnarray}
{\cal A}_{-}+{\cal A}_{+} &=& 8 \pi M^2 G_{N}^2 (1+\alpha) (2+\alpha),\,\, 
{\cal A}_{\pm}- {\cal A}_{\mp} = 8\pi (1+\alpha) M G_{N} T_{\pm}{\cal A}_{\pm}\nonumber\\ 
\frac{1}{{\cal A}_{+}}+\frac{1}{{\cal A}_{-}} &=& \frac{(2+\alpha)}{2\pi M^2 G_{N}^2 (1+\alpha)},\,\,
\frac{1}{{\cal A}_{+}}-\frac{1}{{\cal A}_{-}} =- \frac{1}{\pi M^2 G_{N}^2 \alpha^2\sqrt{1+\alpha} } \nonumber\\ 
T_{+}{\cal S}_{+}+ T_{-}{\cal S}_{-} & =& 0
~.\label{mg1.10}
\end{eqnarray}
The above thermodynamic relations of all the horizons may be used to determine
the MOG BH entropy in terms of the Cardy formula which provides some evidence for a 
BH/CFT correspondence~\cite{castro12}. The most important relation among all the 
above relations is 
\begin{eqnarray}
T_{+}{\cal S}_{+}+ T_{-}{\cal S}_{-} = 0 
\end{eqnarray}
which we may called as a \emph{universal} relation because it indicates that the central charges in left and
right moving sectors of dual CFT in MOG/CFT correspondence are the same for two horizons BH in MOG.  

Moreover, using the above relations we are now able to compute the area bound for this BH following the work of 
Xu et al.~\cite{xu}. Since
$r_{+} \geq r_{-}$  thus ${\cal A}_{+} \geq {\cal A}_{-} \geq 0$.
Now the area product relation gives 
\begin{eqnarray}
{\cal A}_{+} \geq  \sqrt{{\cal A}_{+} {\cal A}_{-}} = 4\pi M^2 G_{N}^2 \alpha (1+\alpha) \geq {\cal A}_{-} ~.\label{mg1.11}
\end{eqnarray}
and the area sum gives
$$
8 \pi M^2 G_{N}^2 (1+\alpha) (2+\alpha)
$$
\begin{eqnarray}
 &=& {\cal A}_{+}+ {\cal A}_{-} \geq {\cal A}_{+}
\geq \frac{{\cal A}_{+}+ {\cal A}_{-}}{2}= 4\pi M^2 G_{N}^2 (1+\alpha) (2+\alpha) \geq {\cal A}_{-}  
~.\label{mg1.12}
\end{eqnarray}
Therefore, one obtains the area bound for  ${\cal H}^{+}$
\begin{eqnarray}
4\pi M^2 G_{N}^2 (1+\alpha) (2+\alpha) \leq {\cal A}_{+} \leq 8 \pi M^2 G_{N}^2 (1+\alpha) (2+\alpha)   ~.\label{mg1.13}
\end{eqnarray}
and  the area bound for  ${\cal H}^{-}$ becomes
\begin{eqnarray}
 0 \leq {\cal A}_{-} \leq 4\pi M^2 G_{N}^2 \alpha (1+\alpha)  ~.\label{mg1.14}
\end{eqnarray}
From these relations one can easily derive the entropy bound for this BH. For completeness, we derive the 
irreducible mass bound for the MOG BH in spherically symmetric cases.
For ${\cal H}^{+}$, it is given by
\begin{eqnarray}
\frac{MG_{N}}{2} \sqrt{(1+\alpha)(2+\alpha)} \leq M_{irr, +} \leq  \frac{MG_{N}}{\sqrt{2}}\sqrt{(1+\alpha)(2+\alpha)} 
~.\label{mg1.15}
\end{eqnarray}
and  for ${\cal H}^{-}$, it is given by
\begin{eqnarray}
0 \leq M_{irr,-} \leq \frac{MG_{N}}{2} \sqrt{\alpha(1+\alpha)}  ~.\label{mg1.16}
\end{eqnarray}

\section{Area Product formula in an axisymmetric MOG BH}
We have seen that in the previous section, the Eq. (\ref{mg1.1}) has the same form as the RN BH solution in the 
Einstein-Maxwell gravity when  ${\cal Q}=\sqrt{\alpha G_{N}}M$ which is just a postulation mentioned in ~\cite{mf7} 
thus one can easily obtain the metric for a Kerr-MOG BH by putting the above unique 
relation in the Kerr-Newman metric~\cite{kn} and the metric~\footnote{It should be noted that the Eq. (\ref{mg1.1})
and Eq. (\ref{mg2.1}) are a new kind of solutions in MOG due to the special unique relation 
${\cal Q}=\sqrt{\alpha G_{N}}M$  in this sense they may be considered as a unique hairy BH solutions in 
the STVG theory.} becomes ~\cite{mf7}
\begin{eqnarray}
ds^2 = \frac{\Delta}{\rho^2} \, \left[dt-a\sin^2\theta d\phi \right]^2-\frac{\sin^2\theta}{\rho^2} \,
\left[(r^2+a^2) \,d\phi-a dt\right]^2
-\rho^2 \, \left[\frac{dr^2}{\Delta}+d\theta^2\right] ~.\label{mg2.1}
\end{eqnarray}
where
\begin{eqnarray}
a &\equiv& \frac{J}{M},\, \rho^2 \equiv r^2+a^2\cos^2\theta \nonumber\\
\Delta &\equiv& r^2-2Mr+a^2 + G_{N}^2 \alpha(1+\alpha) M^2 \equiv(r-r_{+})(r-r_{-})
\end{eqnarray}
It describes the BH with horizon radii:
\begin{eqnarray}
r_{\pm} &=& G_{N} (1+\alpha) M \left[1\pm \sqrt{1-\frac{a^2}{(1+\alpha)^2 M^2 G_{N}^2}-\frac{\alpha}{1+\alpha}}\right]
 ~. \label{mg2.2}
\end{eqnarray}
It should be noted that when $\alpha=0$, one obtains the usual Kerr BH. The above metric is equivalent to the  Kerr-Newman 
BH provided that the gravitational charge ${\cal Q}=\sqrt{\alpha G_{N}}M$.  

The  area of ${\cal H}^\pm$ for the Kerr-MOG BH is
\begin{eqnarray}
{\cal A}_{\pm} &=& 4\pi (1+\alpha) M^2 G_{N}^2 
\left[(2+\alpha)\pm 2(1+\alpha) \sqrt{1-\frac{a^2}{(1+\alpha)^2 M^2 G_{N}^2}-\frac{\alpha}{1+\alpha}} \right] ~.\label{mg2.3}
\end{eqnarray}  
Similarly, the  entropy of ${\cal H}^\pm$ for the Kerr-MOG BH is
\begin{eqnarray}
{\cal S}_{\pm} &=& \pi M^2 G_{N} 
\left[(2+\alpha)\pm 2(1+\alpha) \sqrt{1-\frac{a^2}{(1+\alpha)^2 M^2 G_{N}^2}-\frac{\alpha}{1+\alpha}} \right] ~.\label{mg2.4} 
\end{eqnarray}

The angular velocity of ${\cal H}^\pm$ computed on the horizon is 
\begin{eqnarray}
{\Omega}_{\pm} &=& \frac{a}{(1+\alpha) M^2 G_{N}^2 
\left[(2+\alpha)\pm 2(1+\alpha) \sqrt{1-\frac{a^2}{(1+\alpha)^2 M^2 G_{N}^2}-\frac{\alpha}{1+\alpha}} \right]} ~.\label{mg2.11}
\end{eqnarray}

The Hawking temperature of ${\cal H}^\pm$ should read
\begin{eqnarray}
T_{\pm} &=& \pm \frac{\sqrt{1-\frac{a^2}{(1+\alpha)^2 M^2 G_{N}^2}-\frac{\alpha}{1+\alpha}}}
{2\pi G_{N} M \left[(2+\alpha)\pm 2(1+\alpha) \sqrt{1-\frac{a^2}{(1+\alpha)^2 M^2 G_{N}^2}-\frac{\alpha}{1+\alpha}} \right]}
~.\label{mg2.5} 
\end{eqnarray}

One obtains the Smarr like formula by solving the following quartic equation of $M$:
\begin{eqnarray}
\alpha^2 G_{N}^2 M^4-\left(\frac{\alpha+2}{\alpha+1}\right)\left(\frac{{\cal A}_{\pm}}{2\pi}\right)M^2+4J^2+
\left(\frac{{\cal A}_{\pm}}{4\pi G}\right)^2  &=& 0 .~\label{sm1}
\end{eqnarray}
In the limit $\alpha=0$, one finds the Smarr formula for the Kerr BH. By solving the above Eq. (\ref{sm1}), one obtains 
the Smarr formula for the Kerr-MOG BH:
\begin{eqnarray}
 M^2 &=&  \frac{{\cal A}_{\pm}}{4\pi \alpha^2 G_{N}^2}
 \left[\left(\frac{\alpha+2}{\alpha+1}\right)+\sqrt{\frac{2(\alpha+2)}{(\alpha+1)^2}-\frac{64\pi^2\alpha^2G_{N}^2J^2}
{{\cal A}_{\pm}^2}} \right]~.\label{sm2}
\end{eqnarray}
It can be now easily verified  that Kerr-MOG BH satisfies the first law for both the OH and IH as
\begin{eqnarray}
dM &=& T_{+} d{\cal S}_{+} + \Omega_{+} dJ ~. \label{kmg1}\\
dM &=& -T_{-} d{\cal S}_{-} + \Omega_{-} dJ ~. \label{kmg2}
\end{eqnarray}

The Komar energy for the Kerr-MOG BH is
\begin{eqnarray}
E_{\pm} &=&\pm G_{N} (1+\alpha) M \sqrt{1-\frac{a^2}{(1+\alpha)^2 M^2 G_{N}^2}-\frac{\alpha}{1+\alpha}} .~\label{mg2.6}
\end{eqnarray}
Finally, the Gibbs free energy should read 
$$
G_{\pm} = M \left[1\mp \frac{G_{N} (1+\alpha)}{2}\sqrt{1-\frac{a^2}{(1+\alpha)^2 M^2 G_{N}^2}-
\frac{\alpha}{1+\alpha}}\right]-
$$
\begin{eqnarray}
\left[\frac{a^2}{G_{N}^2 M \left[(2+\alpha)\pm 2(1+\alpha) \sqrt{1-\frac{a^2}{(1+\alpha)^2 M^2 G_{N}^2}-\frac{\alpha}{1+\alpha}} 
\right]}\right] .~\label{mg2.7}
\end{eqnarray}

Now the area product is evaluated to be 
\begin{eqnarray}
{\cal A}_{+} {\cal A}_{-} &=& 64 \pi^2 (1+\alpha)^2 G_{N}^2 \left[J^2+\frac{\alpha^2 G_{N}^2 M^2}{4}\right] 
.~\label{mg2.8}
\end{eqnarray}
Again it is explicitly mass dependent. This means that the area (or entropy) product formula for the Kerr-MOG BH 
is \emph{not} universal \footnote{It might be plausible that the \emph{mass-dependent} formulas that we have derived in 
Eq. (\ref{mg1.9}) and in Eq. (\ref{mg2.8}) do not seem to be \emph{generic} in the STVG theory due to 
the special unique relation ${\cal Q}=\sqrt{\alpha G_{N}}M$. This is just a postulation that has been mentioned in Ref.
~\cite{mf6}. Thus the main resulting \emph{mass-dependence} seems to be an artifact by this postulation. 
Alternatively, we could say that the \emph{mass-dependence} relation may be just the consequence of the unique postulation
${\cal Q}=\sqrt{\alpha G_{N}}M$. Due to this special relation, one could expect that it critically affects  the 
area~(or entropy) formula; that is why the product formula is \emph{not} universal. This is the main difference between the 
Einstein's gravity \& the MOG.}. It follows from the above formula that all 
other thermodynamic products are strictly \emph{mass dependent}. Therefore we could conclude that all the products of 
thermodynamic paramers in MOG are always mass dependent. Thus they could not be treated as a \emph{universal} quantity. 
It could be noted that when the parameter $\alpha$ goes to the zero value, one obtains the above results for Kerr BH. 

\begin{figure}
\begin{center}
{\includegraphics[width=0.45\textwidth]{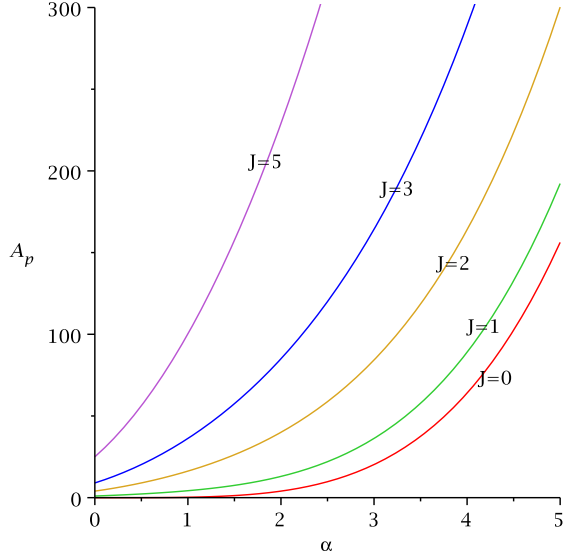}}
\end{center}
\caption{The figure depicts the variation of $A_{p}$  with free parameter $\alpha$ for 
Kerr-MOG BH with $M=G_{N}=1$. Where $A_{p}=\frac{{\cal A}_{+}{\cal A}_{-}}{64 \pi^2}$. \label{mogf}}
\end{figure}

In Fig. \ref{mogf}, we show the axisymmetric MOG BH and  how the area product of 
${\cal H}^{\pm}$ varies with the free parameter. One could observe from the figure due to the presence of the 
spin parameter area product is quite different from that of the spherically symmetric MOG BH.

Proceeding analogously, the irreducible mass product of ${\cal H}^{\pm}$ for the Kerr-MOG BH are
\begin{eqnarray}
M_{irr,+}\, M_{irr,-} &=& \frac{G_{N}(1+\alpha)}{2} \sqrt{J^2+\frac{\alpha^2 G_{N}^2 M^2}{4}} ~. \label{mg2.9}
\end{eqnarray}

The other important thermodynamic relations for Kerr-MOG BH are
\begin{eqnarray}
{\cal A}_{-}+{\cal A}_{+} &=& 8 \pi M^2 G_{N}^2 (1+\alpha) (2+\alpha) ,\,\,
{\cal A}_{\pm}- {\cal A}_{\mp} = 8\pi (1+\alpha) M G_{N} T_{\pm} {\cal A}_{\pm} \nonumber\\
\frac{1}{{\cal A}_{+}}+\frac{1}{{\cal A}_{-}} &=& \frac{M^2(2+\alpha)}{8\pi (1+\alpha)
\left[J^2+\frac{\alpha^2 G_{N}^2 M^2}{4}\right]} \nonumber\\
\frac{1}{{\cal A}_{+}}-\frac{1}{{\cal A}_{-}} &=& -\frac{M\sqrt{G_{N}^2M^2(1+\alpha)-a^2}}
{4\pi G_{N}(1+\alpha)\left[J^2+\frac{\alpha^2 G_{N}^2 M^2}{4}\right] } \nonumber\\
 T_{+}{\cal S}_{+}+ T_{-}{\cal S}_{-} &=& 0 ,\,\, \frac{{\Omega}_{+}}{T_{+}}+\frac{{\Omega}_{-}}{T_{-}} = 0
~.\label{mg2.10}
\end{eqnarray}
These relations for all the horizons of the Kerr-MOG BH could be further used to determine
the BH entropy in terms of the Cardy formula which gives some clue for a MOG BH/CFT 
correspondence in the axisymmetric spacetime. The following \emph{universal} relation 
\begin{eqnarray}
T_{+}{\cal S}_{+}+ T_{-}{\cal S}_{-} = 0 
\end{eqnarray}
implies that the central charges in the left and right moving modes of the dual CFT in MOG/CFT 
correspondence are same for the MOG-Kerr BH.

It should be mentioned that in ~\cite{chen12,chen13},``the area product being mass-independent is equivalent to the relation 
$T_{+}{\cal S}_{+}= T_{-}{\cal S}_{-}$''. We proved that this argument is violated in the Kerr-MOG BH, since here, we see 
that the relation $T_{+}{\cal S}_{+}= -T_{-}{\cal S}_{-}$ is satisfied but the area product is \emph{not} mass-independent.
This is an interesting observation for MOG.

Analogously, using the above relations one can easily calculate the area bound for this BH. Since
$r_{+} \geq r_{-}$  thus ${\cal A}_{+} \geq {\cal A}_{-} \geq 0$.
Now the area product relation is given by 
\begin{eqnarray}
{\cal A}_{+} \geq  \sqrt{{\cal A}_{+} {\cal A}_{-}}= 8 \pi G_{N} (1+\alpha) \sqrt{J^2+\frac{\alpha^2 G_{N}^2 M^2}{4}} 
\geq {\cal A}_{-} ~.\label{m2.11}
\end{eqnarray}
and the area sum is calculated to be
$$
8 \pi M^2 G_{N}^2 (1+\alpha) (2+\alpha)
$$
\begin{eqnarray}
 &=& {\cal A}_{+}+ {\cal A}_{-} \geq {\cal A}_{+}
\geq \frac{{\cal A}_{+}+ {\cal A}_{-}}{2}= 4\pi M^2 G_{N}^2 (1+\alpha) (2+\alpha) \geq {\cal A}_{-}  
~.\label{mg2.12}
\end{eqnarray}
Likewise, the area bound for  ${\cal H}^{+}$ is
\begin{eqnarray}
4\pi M^2 G_{N}^2 (1+\alpha) (2+\alpha) \leq {\cal A}_{+} \leq 8 \pi M^2 G_{N}^2 (1+\alpha) (2+\alpha) ~.\label{mg2.13}
\end{eqnarray}
and  the area bound for  ${\cal H}^{-}$ is
\begin{eqnarray}
 0 \leq {\cal A}_{-} \leq 8 \pi G_{N} (1+\alpha) \sqrt{J^2+\frac{\alpha^2 G_{N}^2 M^2}{4}}   ~.\label{mg2.14}
\end{eqnarray}
From these relations,  one could easily derive the entropy bound for this BH. It should be noted that, in the limit 
$\alpha=0$, one obtains the result for the Kerr BH~\cite{xu}.

For our record, one could derive the irreducible mass bound for MOG-Kerr BH.
Thus for ${\cal H}^{+}$, it is calculated to be
\begin{eqnarray}
\frac{MG_{N}}{2} \sqrt{(1+\alpha)(2+\alpha)} \leq M_{irr, +} \leq  \frac{MG_{N}}{\sqrt{2}}\sqrt{(1+\alpha)(2+\alpha)} 
~.\label{mg2.15}
\end{eqnarray}
and  for ${\cal H}^{-}$, it is 
\begin{eqnarray}
0 \leq M_{irr,-} \leq \sqrt{\frac{G_{N}(1+\alpha)}{2}} \left(J^2+\frac{\alpha^2 G_{N}^2 M^2}{4}\right)^\frac{1}{4}  
~.\label{mg2.16}
\end{eqnarray}

Finally, it should be noted that using the symmetric properties of $r_{\pm}$, one obtains 
\begin{eqnarray}
T_{-} &=& -T_{+}|_{r_{+}\leftrightarrow r_{-}},\, {\cal S}_{-}={\cal S}_{+}|_{r_{+}\leftrightarrow r_{-}}, \,
\Omega_{-}=\Omega_{+}|_{r_{+}\leftrightarrow r_{-}},\, {\cal A}_{-}={\cal A}_{+}|_{r_{+}\leftrightarrow r_{-}}, \nonumber\\
M_{irr, -} &=& -M_{irr, +}|_{r_{+}\leftrightarrow r_{-}},\, 
E_{-} = -E_{+}|_{r_{+}\leftrightarrow r_{-}},\,
T_{-}{\cal S}_{-}=-T_{+}{\cal S}_{+}|_{r_{+}\leftrightarrow r_{-}}
~.\label{mg2.18}
\end{eqnarray} 

Perhaps, most importantly,  the \emph{mass-dependent} formulas that we have derived in Eq. (\ref{mg1.9}) and 
in Eq. (\ref{mg2.8}) do not seem to be \emph{generic} in MOG, since the metrics of Eq. \ref{mg1.1} 
and Eq. \ref{mg2.1} have the same form as the RN BH solution and the KN BH solution in the Einstein-Maxwell system
~\footnote{It should be mentioned that MOG is a purely gravitional theory depending on only mass and spin. The metric 
solutions are similar algebraically to RN metrics and KN metrics but this comparison should end there. Because astrophysical 
bodies and BHs are electrically neutral. The electric charge if present is negligible and a BH would instantly 
blow up if $Q_{electric}=M$ due to the ratio of Coulomb force to gravity~$10^{40}$. Any electric charge would have 
very little effect on the spacetime metric.}. 
Due to the special unique relation ${\cal Q}=\sqrt{\alpha G_{N}}M$ which is a postulation we have mentioned 
earlier,  the main \emph{mass-dependence} result seems to just an artifact due this postulation. In the modified 
theory of gravity, there are several examples of BHs where the charge associated with the new degrees of freedom ~
(scalar or vector field) is related to the mass, whereas the charge is secondary. An example of this type is the  
very  controversial Bocharova-Bronnikov-Melnikov-Bekenstein~(BBMB) BH~\cite{bbm} solution in Einstein-conformally 
coupled scalar theory, where electric and magnetic fields of Einstein-Maxwell systems relate $Q=M$ and 
consequently the metric is obtained as extreme RN BH solution. Also this is a single parameter solutions where 
the parameter is $M$, only the total mass parameter with the scalar field which is diverges at the horizon but the 
geometry is regular there. But in our case, the metrics of Eq. \ref{mg1.1} and Eq. \ref{mg2.1} do not not seem 
to be this type, more importantly the charge parameter ${\cal Q}$ is quite independent from the mass parameter
$M$. Hence the \emph{mass-dependence} relation might be just the consequences of the unique postulation
${\cal Q}=\sqrt{\alpha G_{N}}M$.

\section{Kerr-MOG/CFT correspondence}
In this section, we should derive the central charges $c_{R}$ and $c_{L}$ of the right and left moving sectors 
of the dual CFT in Kerr-MOG/CFT correspondence. We should prove that the central charges of the right and left moving 
sectors are same i.e. $c_{R}=c_{L}$ for Kerr-MOG BH. Also we should determine the dimensionless  temperature of 
microscopic CFT from the above thermodynamic relations. Furthermore using Cardy formula,  we should explicitly compute 
the right and left moving entropies in 2D CFT. Moreover, in the extreme limit we find the Frolov-Thorne vacuum state 
temperature which is thermally populated with a Boltzmann distribution. Finally by using Cardy formula we should determine 
the  microscopic entropy of extreme Kerr-MOG BH and it is exactly same to the macroscopic Bekenstein-Hawking entropy .

Now we could proceed as in terms of OH radius $r_{+}$ and IH radius $r_{-}$, we could write the ADM mass and spin parameter  
\begin{eqnarray}
M=\frac{\left(r_{+}+ r_{-}\right)}{2G}  \,\, \mbox{and}\,\, a=\sqrt{r_{+} r_{-}-\left(\frac{\alpha}{1+\alpha}\right)
\frac{(r_{+}+ r_{-})^2}{4}}  ~.\label{km1}
\end{eqnarray}

Therefore the angular momentum is derived to be 
\begin{eqnarray}
J=\frac{\left(r_{+}+ r_{-}\right)}{2G}\sqrt{r_{+} r_{-}-\left(\frac{\alpha}{1+\alpha}\right)\frac{(r_{+}+ r_{-})^2}{4}}   
~.\label{km2}
\end{eqnarray}

Again in terms of $r_{+}$ and $r_{-}$, we could derive the entropy, Hawking temperature and angular velocity
 of  ${\cal H}^{+}$ for Kerr-MOG BH
\begin{eqnarray}
S_{+} &=& \frac{\pi \left(r_{+}+ r_{-}\right)}{2G}\left[2r_{+}-\left(\frac{\alpha}{1+\alpha}\right)\frac{(r_{+}+ r_{-})}{2} 
\right]  ~.\label{km3}\\
T_{+} &=& \frac{(r_{+}-r_{-})}{2\pi (r_{+}+ r_{-})\left[2r_{+}-\left(\frac{\alpha}{1+\alpha}\right)\frac{(r_{+}+ r_{-})}{2} 
\right] } ~.\label{km4}\\
\Omega_{+} &=& \frac{\sqrt{r_{+} r_{-}-\left(\frac{\alpha}{1+\alpha}\right)
\frac{(r_{+}+ r_{-})}{2}}}{\frac{\left(r_{+}+ r_{-}\right)}{2}\left[2r_{+}-\left(\frac{\alpha}{1+\alpha}\right)\frac{(r_{+}+ r_{-})}{2} 
\right]}  ~.\label{km5}
\end{eqnarray}
Using the features of symmetry of $r_{\pm}$, one obtains the following relation for ${\cal H}^{-}$
\begin{eqnarray}
T_{-} &=& -T_{+}|_{r_{+}\leftrightarrow r_{-}}, S_{-}=S_{+}|_{r_{+}\leftrightarrow r_{-}}, 
\Omega_{-}=\Omega_{+}|_{r_{+}\leftrightarrow r_{-}}
~.\label{km7}
\end{eqnarray}
Thus the first law of BH thermodynamics may be rewritten as in terms of right and left moving sectors
of dual CFT.
\begin{eqnarray}
\frac{dM}{2} &=& T_{R} dS_{R}+\Omega_{R} dJ    ~.\label{km8} \\
             &=& T_{L} dS_{L}+\Omega_{L} dJ    ~.\label{km9}
\end{eqnarray}
and using the definitions of $\beta_{R,L}=\beta_{+}\pm \beta_{-}$, $\beta_{\pm}=\frac{1}{T_{\pm}}$, 
$\Omega_{R,L}=\frac{\beta_{+}\Omega_{+}\pm \beta_{-}\Omega_{-}}{2 \beta{R,L}}$,  
and $S_{R,L}=\frac{(S_{+}\mp S_{-})}{2}$~\cite{cv96,cv97,cvf97}. 

Now one could easily derive the temperature and entropy for left moving sectors and right moving sectors as
\begin{eqnarray}
T_{L} &=& \frac{1}{4\pi \left(r_{+}+ r_{-} \right)}, \,\,\, 
T_{R} = \left(\frac{\alpha+1}{\alpha+2} \right)\frac{(r_{+}-r_{-})}{2\pi (r_{+}+r_{-})^2 } \nonumber\\
S_{L} &=& \left(\frac{\alpha+2}{\alpha+1} \right) \frac{\pi (r_{+}+r_{-})^2}{4G} , \,\,\, 
S_{R} =  \frac{\pi (r_{+}^2-r_{-}^2)}{2G} \nonumber\\
\Omega_{L} &=& 0,  \,\,\, \Omega_{R} = 2\left(\frac{\alpha+1}{\alpha+2}\right) 
\frac{\sqrt{r_{+} r_{-}-\left(\frac{\alpha}{1+\alpha}\right)
\frac{(r_{+}+ r_{-})^2}{4}}}{(r_{+}+r_{-})^2}
\nonumber\\
~.\label{km10}
\end{eqnarray}
With the help of Eq. (\ref{km8}) \& Eq. (\ref{km9}), one could obtain the first law of BH thermodynamics for 
left moving sectors and right moving sectors of dual CFT
\begin{eqnarray}
dJ &=& \frac{T_{L}}{\Omega_{R}-\Omega_{L}} dS_{L}- \frac{T_{R}}{\Omega_{R}-\Omega_{L}} dS_{R} ~.\label{km11}
\end{eqnarray}
From the above Eq. (\ref{km11}), one could easily determine the  dimensionless temperature of the left and right
moving sectors of the dual CFT correspondence. They are defined as  
\begin{eqnarray}
T_{L}^{J} &=& \frac{T_{L}}{\Omega_{R}-\Omega_{L}}, \,\, T_{R}^{J} = \frac{T_{R}}{\Omega_{R}-\Omega_{L}}  ~.\label{km12}
\end{eqnarray}
For Kerr-MOG BH, one can derived to be
\begin{eqnarray}
T_{L}^{J} &=& \left(\frac{\alpha+2}{\alpha+1}\right) \frac{(r_{+}+r_{-})}
{8\pi \sqrt{r_{+} r_{-}-\left(\frac{\alpha}{1+\alpha}\right) \frac{(r_{+}+ r_{-})^2}{4}}}    ~.\label{km13}
\end{eqnarray}
\&
\begin{eqnarray}
T_{R}^{J} &=&  \frac{(r_{+}-r_{-})}
{4\pi \sqrt{r_{+} r_{-}-\left(\frac{\alpha}{1+\alpha}\right) \frac{(r_{+}+ r_{-})^2}{4}}}    ~.\label{km14}
\end{eqnarray}
they are exactly the microscopic temperature of dual CFT

Now we are ready to determine the central charges~\cite{chen12} in left and right moving sectors of the Kerr-MOG/CFT 
correspondence via the Cardy formula 
\begin{eqnarray}
S_{L}^{J} &=& \frac{\pi^2}{3}c_{L}^{J}T_{L}^{J},\,\,\, S_{R}^{J} = \frac{\pi^2}{3}c_{R}^{J}T_{R}^{J}~.\label{kd14}
\end{eqnarray}
Thus the central charges of dual CFT are
\begin{eqnarray}
c_{L}^{J} &=& 12J,\,\,\,  c_{R}^{J}=12J    ~.\label{km15}
\end{eqnarray}
This means that the central charges of left moving sectors and right moving sectors of dual CFT are same for 
Kerr-MOG BH. This is an interesting result for \emph{Kerr MOG BH}. It is also more interesting because it is 
independent of free parameter $\alpha$ \& the result is exactly same as we have seen in case of  Kerr BH~\cite{hartman9} 
and Kerr-Newman BH~\cite{chen12}. This kind of observation indicates that Kerr-MOG BH is dual to  
$c_{L}=c_{R}=12J$ of 2D CFT at temperature $(T_{L},T_{R})$ for each value of $M$ and $J$.

Now we are going to see what happens in the extremal limit $r_{+}=r_{-}$? 
\begin{eqnarray}
T_{L} &=& \frac{1}{8\pi r_{+}}, \,\,\,  T_{R} = 0 \nonumber\\
S_{L} &=&  \left(\frac{\alpha+2}{\alpha+1}\right) \frac{\pi r_{+}^2}{G}  , \,\,\, 
S_{R} = 0 \nonumber\\
\Omega_{L} &=& 0,  \,\,\, \Omega_{R} =\frac{\sqrt{1+\alpha}}{2+\alpha} \frac{1}{2 r_{+}^2} ~.\label{km16}
\end{eqnarray}
\begin{eqnarray}
T_{L}^{J} &=&  \frac{1}{4\pi} \frac{\alpha+2}{\sqrt{1+\alpha}}, \,\,\,  T_{R}^{J} = 0  ~.\label{km17}
\end{eqnarray}
this is the left moving temperature which is actually Frolov-Thorne vacuum quantum state temperature~\cite{frolov89}, and 
finally one obtains the central charge for extremal Kerr-MOG BH
\begin{eqnarray}
c_{L}^{J} &=& 12J ~.\label{km18}
\end{eqnarray}
Finally, one could obtain the microscopic entropy via Cardy formula in chiral dual CFT
\begin{eqnarray}
S_{micro} &=& \frac{\pi^2}{3}c_{L}^{J}T_{L}^{J}= \frac{\alpha+2}{\sqrt{1+\alpha}} \pi J    ~.\label{km19}
\end{eqnarray}
which is perfectly match with the macroscopic Bekenstein-Hawking entropy for extreme Kerr-MOG BH. In the 
limit $\alpha=0$, one finds the macroscopic Bekenstein-Hawking entropy for extreme Kerr BH, $S_{micro}=2\pi J$ 
~\cite{guica}. 

In the following section, we shall analyze the thermodynamics properties of regular BH in MOG which is so called singularity 
free solution of classical general theory of relativity. First, the idea of regular BH solution has been incorporated by 
Bardeen in 1980~\cite{bard}. Subsequently, Ay\'{o}n-Beato and Garc\'{i}a~\cite{abg} derived a singularity-free solution 
of the Einstein field equations which is coupled to a non-linear electrodynamics  in 1998.

\section{Area Product formula in MOG regular BH}
The metric of a regular MOG BH~\cite{mf7} is given by 
\begin{eqnarray}
ds^2=-{\cal U}(r)dt^{2}+\frac{dr^{2}}{{\cal U}(r)}+r^{2}(d\theta^{2}+\sin^{2}\theta d\phi^{2}) ~.\label{rm3.1}
\end{eqnarray}
where the function ${\cal U}(r)$ is defined by
\begin{eqnarray}
{\cal U}(r) &=& 1-\frac{2GMr^2}{(r^2+\alpha G_{N}GM^2)^{\frac{3}{2}}}+\frac{\alpha G_{N}GM^2r^2}{(r^2+\alpha G_{N}GM^2)^2} 
~.\label{rm3.2}
\end{eqnarray}
This is a solution with an EH but without singularities. It should be noted that if the value of the free parameter 
$\alpha$ is less than a critical value i.e. $\alpha_{c}=0.673$ then there exists two physical horizons namely EH \&
CH. Otherwise for $\alpha>\alpha_{c}$, there exists no horizon~\cite{mf7}.

To obtain the BH horizon equation we have to set  ${\cal U}(r)=0$ i.e.
\begin{equation}
r^{8}+(6\alpha G_{N}GM^2 -4G^2M^2)r^{6}+(11\alpha^2 G_{N}^2 G^2M^4-4\alpha G_{N}G^3M^4)r^{4} 
$$ $$
+6(\alpha G_{N}GM^2)^3r^2+(\alpha G_{N}GM^2)^4 = 0 ~.\label{rm3.3}
\end{equation}
Now putting $r^2=x$, one obtains the following polynomial equation
\begin{equation}
x^{4}+(6\alpha G_{N}GM^2 -4G^2M^2)x^{3}+(11\alpha^2 G_{N}^2 G^2M^4-4\alpha G_{N}G^3M^4)x^{2} 
$$ $$
+6(\alpha G_{N}GM^2)^3 x+(\alpha G_{N}GM^2)^4 = 0 ~.\label{rm3.4}
\end{equation}
This is a fourth order polynomial equation. To finding the roots one could apply the Vieta's theorem and one obtains
\begin{eqnarray}
\sum_{i=1}^{4} x_{i} &=& 4G^2M^2-6\alpha G_{N}GM^2 ~,\label{eq1}\\
\sum_{1\leq i<j\leq 4} x_{i}x_{j} &=& 11\alpha^2 G_{N}^2 G^2M^4-4\alpha G_{N}G^3M^4  ~,\label{eq2}\\
\sum_{1\leq i<j<k\leq 4} x_{i}x_{j} x_{k} &=& -6(\alpha G_{N}GM^2)^3 ~,\label{eq3}\\
\prod_{i=1}^{4}x_{i} &=& (\alpha G_{N}GM^2)^4 ~.\label{eq4}
\end{eqnarray} 
Eliminating the mass parameter, we should find the following mass-independent equation
\begin{eqnarray}
\sum_{1\leq i<j\leq 4} x_{i}x_{j} &=& \frac{(7\alpha^2-4\alpha)}{(4-2\alpha)^2}  \left(\sum_{i=1}^{4} x_{i} \right)^2  ~,\label{eq5}\\
\sum_{1\leq i<j<k\leq 4} x_{i}x_{j} x_{k} &=& \frac{6\alpha^3}{(2\alpha-4)^3} \left(\sum_{i=1}^{4} x_{i} \right)^3 ~,\label{eq6}\\
\prod_{i=1}^{4}x_{i} &=& \frac{\alpha^4}{(4-2\alpha)^4} \left(\sum_{i=1}^{4} x_{i} \right)^4  ~.\label{eq7}
\end{eqnarray} 
In terms of area ${\cal A}_{i}=4 \pi x_{i}$, the above mass-independent equation could be written as 
\begin{eqnarray}
\sum_{1\leq i<j\leq 4} {\cal A}_{i}{\cal A}_{j} &=&\frac{(7\alpha^2-4\alpha)}{(4-2\alpha)^2} 
\left(\sum_{i=1}^{4} {\cal A}_{i}\right)^2  ~,\label{eq9}\\
\sum_{1\leq i<j<k\leq 4} {\cal A}_{i}{\cal A}_{j} {\cal A}_{k} &=& \frac{6\alpha^3}{(2\alpha-4)^3} 
\left(\sum_{i=1}^{4} {\cal A}_{i}\right)^3 ~,\label{eq10}\\
\prod_{i=1}^{4}{\cal A}_{i} &=& \frac{\alpha^4}{(4-2\alpha)^4} \left(\sum_{i=1}^{4} {\cal A}_{i}\right)^4 ~.\label{eq11}
\end{eqnarray} 
Eliminating further one obtains the following mass independent relation 
\begin{eqnarray}
\sum_{1\leq i<j<k\leq 4} {\cal A}_{i}{\cal A}_{j} {\cal A}_{k} &=& \left(\frac{6\alpha}{2\alpha-4}\right)^\frac{3}{2} 
\left(\sum_{1\leq i<j\leq 4} {\cal A}_{i}{\cal A}_{j}\right)^\frac{3}{2} ~,\label{eq12}\\
\prod_{i=1}^{4}{\cal A}_{i} &=& \left(\frac{\alpha}{7\alpha-2}\right)^2 
\left(\sum_{1\leq i<j\leq 4} {\cal A}_{i}{\cal A}_{j}\right)^2 ~.\label{eq13}
\end{eqnarray} 
It must be noted that the above formulae for the horizon areas could be obtained by observing that the horizon coordinate 
positions $r_{i}$ are related to the zeros $x_{i}$ of a fourth order polynomial via the above relation $r_{i}^2=x$.
On the other hand any horizon must indeed be a zero of that polynomial, it is by no means clear that any zero also 
corresponds to a horizon. Indeed, for a physically relevant horizon, the zero must be real and positive. Therefore, 
equations that involve all four roots of the polynomial are not necessarily physically resonable. This is depends 
upon the value of free parameter $\alpha$ that has been discussed above. Hence, there are at most two actual horizons 
so that the above equations that we have derived which depend on all four roots, are all unphysical. Reasonable 
equations could be constructed if two roots are eliminated such that relations between only two horizon areas are 
obtained. 

One could write the explicit expressions in terms of two horizons area  by using the equations 
~(\ref{eq1}), (\ref{eq2}), (\ref{eq3}) and (\ref{eq4}). Thus one obtains 
\begin{equation}
\frac{{\cal A}_{1}{\cal A}_{2}}{16 \pi^2}-\frac{({\cal A}_{1}+{\cal A}_{2})^2}{16 \pi^2}=
\left(11\alpha^2 G_{N}^2 G^2-4\alpha G_{N}G^3\right)M^4-
$$ $$ 
\left(\frac{{\cal A}_{1}+{\cal A}_{2}}{4\pi}\right) 
(4G^2-6\alpha G_{N}G)M^2-\frac{16\pi^2(\alpha G_{N}G)^4M^8}{{\cal A}_{1}{\cal A}_{2}}
~.\label{13a}
\end{equation}
and
\begin{equation}
 \frac{{\cal A}_{1}{\cal A}_{2}}{64\pi^3}= \left(\frac{{\cal A}_{1}{\cal A}_{2}}{4\pi}\right)(4G^2-6\alpha G_{N}G)M^2+
 $$ $$
4\pi (\alpha G_{N}G)^4\left(\frac{{\cal A}_{1}+{\cal A}_{2}}{4\pi}\right)M^8
+6(\alpha G_{N}G)^3M^6 ~.\label{13b}
\end{equation}
It could be easily seen that from the above two equations, there has been no way to eliminate 
the mass parameter from these equations. It indicates that even for regular MOG BH there has been no way to construct 
mass-independent formula in terms of two physical horizons area. It is indeed true that the mass-independent relations 
that we have constructed in terms of four horizons area which are unphysical. This is also an another new result 
for regular MOG BH.

The Hawking temperature of ${\cal H}^\pm$ for MOG regular BH should read 
\begin{eqnarray}
T_{\pm} &=& \frac{GM}{2\pi}\left[\frac{r_{\pm}\left(r_{\pm}^2-2\alpha G_{N}GM^2 \right)}
{\left(r_{\pm}^2+\alpha G_{N}GM^2 \right)^\frac{5}{2}}\right]-\frac{\alpha G_{N}GM^2}{2\pi}
\left[\frac{r_{\pm}\left(r_{\pm}^2-\alpha G_{N}GM^2 \right)}
{\left(r_{\pm}^2+\alpha G_{N}GM^2 \right)^3}\right] ~.\label{eq14}~\nonumber
\end{eqnarray}
Whereas the Komar energy of ${\cal H}^\pm$ is derived to be 
\begin{eqnarray}
E_{\pm} &=& GM \left[\frac{r_{\pm}^3\left(r_{\pm}^2-2\alpha G_{N}GM^2 \right)}
{\left(r_{\pm}^2+\alpha G_{N}GM^2 \right)^\frac{5}{2}}\right]-\alpha G_{N}G M^2 
\left[\frac{r_{\pm}^3\left(r_{\pm}^2-\alpha G_{N}GM^2 \right)}
{\left(r_{\pm}^2+\alpha G_{N}GM^2 \right)^3}\right].~\label{eq15}~\nonumber
\end{eqnarray}
where $r_{\pm}$ is the root of Eq. (\ref{rm3.3}).
Finally, the Gibbs free energy is computed to be 
\begin{eqnarray}
G_{\pm} &=&M- \frac{GM}{2} \left[\frac{r_{\pm}^3\left(r_{\pm}^2-2\alpha G_{N}GM^2 \right)}
{\left(r_{\pm}^2+\alpha G_{N}GM^2 \right)^\frac{5}{2}}\right]-\frac{\alpha G_{N}G M^2}{2} 
\left[\frac{r_{\pm}^3\left(r_{\pm}^2-\alpha G_{N}GM^2 \right)}
{\left(r_{\pm}^2+\alpha G_{N}GM^2 \right)^3}\right].~\label{eq16}~\nonumber
\end{eqnarray}
From the above thermodynamic relations, one can conclude that the product is strictly mass dependent.

\section{\label{dis} Conclusion}
We investigated the features of \emph{inner} and outer horizon thermodynamics of MOG. We derived the thermodynamic product 
relations particularly emphasized on area~(or entropy) products for this gravity. We considered both spherically symmetric 
solution and axisymmetric solution of MOG. We found that the area (or entropy) product formula for both cases is 
\emph{not} mass-independent  because they depends on ADM parameter while in EG this formula is universal.  
We also examined  the \emph{first law} which is fulfilled  at the IH as well as OH. We also derived other thermodynamic 
relations like products and sums. 

We  further derived the \emph{Smarr mass formula} and 
\emph{Christodoulou's irreducible mass formula} for this kind of BH 
in MOG.  Moreover, we derived  the area (or  entropy) bound for all the horizons. Furthermore, we showed the central 
charges of the left and right moving modes of the dual CFT in MOG/CFT correspondence  are same by using a universal 
thermodynamic relations. For regular MOG BH, we derived some complicated combinations of four horizon area (or entropy) 
product relations that are mass independent but it is unphysical. On the other hand, we derived explicit expressions 
in terms of two physical horizons area while it is mass dependent.

We proved that the statement could made in ~\cite{chen12,chen13},``the area product being mass-independent is equivalent 
to the relation $T_{+}{\cal S}_{+}= T_{-}{\cal S}_{-}$'' breaks down in case of Kerr-MOG BH. Interestingly,  we pointed out 
that the relation $T_{+}{\cal S}_{+}= -T_{-}{\cal S}_{-}$ is satisfied but the area product is \emph{not} mass-independent.
Moreover, we derived the central charges for Kerr-MOG BH, $c_{L}=12J$ which is usually derived using asymptotic 
symmetry group analysis. We also computed the dimensionless temperature for extreme Kerr-MOG BH. Using famous Cardy formula, we 
derived the microscopic entropy for extreme Kerr-MOG BH which is precisely equal to the macroscopic Bekenstein-Hawking entropy. 
Thus we conjectured that extreme Kerr-MOG BH is holographically dual to a chiral 2D CFT with $c_{L}=12J$.

To sum up, the Ansorg \& Hennig's  ``mass-independence conjecture'' could break down in case of MOG due to the special unique 
relation ${\cal Q}=\sqrt{\alpha G_{N}}M$ consequently it critically affects on the thermodynamic product relations.

\section*{Acknowledgements}
I am grateful to Prof. John Moffat of Perimeter Institute, Waterloo, Ontario, Canada for reading the manuscript 
carefully and giving me the valuable suggestions. I also have been benefited for the suggestions from Prof. Sumit R. Das 
of Kentucky University during the Poster presentations at KIAS-YITP joint workshop 2017 
``Strings, Gravity and Cosmology"~(2017/09/19 --- 2017/09/22 ) held at YITP, Kyoto University, Japan. 

This work was supported in part by the  Yukawa Institute for Theoretical Physics, Kyoto University~(YITP) 
during a visit for participating in the program KIAS-YITP joint workshop~2017 ``Strings, Gravity and Cosmology"~
(Workshop Number:YITP-W-17-12).
\begin{center}
 \&
\end{center}
This research was also supported in part by the International Centre for Theoretical Sciences~(ICTS) 
during a visit for participating in the program ``US-India Advanced Studies Institute: 
Classical and Quantum Information~(Code: ICTS/Prog-infoasi/2016/12)''.

\end{document}